\documentclass [12pt] {article}
\usepackage{a4}

\usepackage{amsfonts}
\usepackage{amsmath}

  \def\pp{{\mathchoice
          {
              \kern 1pt%
              \raise 1pt
              \vbox{\hrule width5pt height0.4pt depth0pt
                    \kern -2pt
                    \hbox{\kern 2.3pt
                          \vrule width0.4pt height6pt depth0pt
                          }
                    \kern -2pt
                    \hrule width5pt height0.4pt depth0pt}%
                    \kern 1pt
           }
            {
              \kern 1pt%
              \raise 1pt
              \vbox{\hrule width4.3pt height0.4pt depth0pt
                    \kern -1.8pt
                    \hbox{\kern 1.95pt
                          \vrule width0.4pt height5.4pt depth0pt
                          }
                    \kern -1.8pt
                    \hrule width4.3pt height0.4pt depth0pt}%
                    \kern 1pt
            }
            {
              \kern 0.5pt%
              \raise 1pt
              \vbox{\hrule width4.0pt height0.3pt depth0pt
                    \kern -1.9pt  
                    \hbox{\kern 1.85pt
                          \vrule width0.3pt height5.7pt depth0pt
                          }
                    \kern -1.9pt
                    \hrule width4.0pt height0.3pt depth0pt}%
                    \kern 0.5pt
            }
            {
              \kern 0.5pt%
              \raise 1pt
              \vbox{\hrule width3.6pt height0.3pt depth0pt
                    \kern -1.5pt
                    \hbox{\kern 1.65pt
                          \vrule width0.3pt height4.5pt depth0pt
                          }
                    \kern -1.5pt
                    \hrule width3.6pt height0.3pt depth0pt}%
                    \kern 0.5pt
            }
        }}

  \def\mm{{\mathchoice
                       {
                             \kern 1pt
               \raise 1pt    \vbox{\hrule width5pt height0.4pt depth0pt
                                  \kern 2pt
                                  \hrule width5pt height0.4pt depth0pt}
                             \kern 1pt}
                       {
                            \kern 1pt
               \raise 1pt \vbox{\hrule width4.3pt height0.4pt depth0pt
                                  \kern 1.8pt
                                  \hrule width4.3pt height0.4pt depth0pt}
                             \kern 1pt}
                       {
                            \kern 0.5pt
               \raise 1pt
                            \vbox{\hrule width4.0pt height0.3pt depth0pt
                                  \kern 1.9pt
                                  \hrule width4.0pt height0.3pt depth0pt}
                            \kern 1pt}
                       {
                           \kern 0.5pt
             \raise 1pt  \vbox{\hrule width3.6pt height0.3pt depth0pt
                                  \kern 1.5pt
                                  \hrule width3.6pt height0.3pt depth0pt}
                           \kern 0.5pt}
                       }}

\catcode`@=11
\def\un#1{\relax\ifmmode\@@underline#1\else
        $\@@underline{\hbox{#1}}$\relax\fi}
\catcode`@=12

\let\du=\du                     

\def\c{\chi}
\def\d{\delta}
\def\e{\epsilon}

\def\g{\gamma}

\def\j{\psi}
\def\k{\kappa}
\def\l{\lambda}
\def\m{\mu}
\def\n{\nu}
\def\o{\omega}

\def\q{\theta}
\def\r{\rho}
\def\s{\sigma}

\def\cl{{\cal L}}

\def\bo{{\raise-.3ex\hbox{\large$\Box$}}}               
\def\pa{\partial}                                       
\def\TH{{\raise.2ex\hbox{$\displaystyle \bigodot$}\mskip-4.7mu \llap H \;}}
\def\face{{\raise.2ex\hbox{$\displaystyle \bigodot$}\mskip-2.2mu \llap {$\ddot
        \smile$}}}                                      

\def\sp#1{{}^{#1}}                              
\def\sb#1{{}_{#1}}                              
\def\leftrightarrowfill{$\mathsurround=0pt \mathord\leftarrow \mkern-6mu
        \cleaders\hbox{$\mkern-2mu \mathord- \mkern-2mu$}\hfill
        \mkern-6mu \mathord\rightarrow$}
\def\dvec#1{\vbox{\ialign{##\crcr
        \leftrightarrowfill\crcr\noalign{\kern-1pt\nointerlineskip}
        $\hfil\displaystyle{#1}\hfil$\crcr}}}           

\def\frac#1#2{{\textstyle{#1\over\vphantom2\smash{\raise.20ex
        \hbox{$\scriptstyle{#2}$}}}}}                   
\def\sfrac#1#2{{\vphantom1\smash{\lower.5ex\hbox{\small$#1$}}\over
        \vphantom1\smash{\raise.4ex\hbox{\small$#2$}}}} 
\def\bfrac#1#2{{\vphantom1\smash{\lower.5ex\hbox{$#1$}}\over
        \vphantom1\smash{\raise.3ex\hbox{$#2$}}}}       
\def\afrac#1#2{{\vphantom1\smash{\lower.5ex\hbox{$#1$}}\over#2}}    

\def\[{\lfloor{\hskip 0.35pt}\!\!\!\lceil}
\def\]{\rfloor{\hskip 0.35pt}\!\!\!\rceil}

\def\du#1#2{_{#1}{}^{#2}}
\def\ud#1#2{^{#1}{}_{#2}}

\def\ha{{\fracmm12}}

\def\un{\underline}
\def\fracmm#1#2{{{#1}\over{#2}}}

\def\low#1{{\raise -3pt\hbox{${\hskip 0.75pt}\!_{#1}$}}}

\newskip\humongous \humongous=0pt plus 1000pt minus 1000pt
\def\caja{\mathsurround=0pt}
\def\eqalign#1{\,\vcenter{\openup2\jot \caja
        \ialign{\strut \hfil$\displaystyle{##}$&$
        \displaystyle{{}##}$\hfil\crcr#1\crcr}}\,}
\newif\ifdtup

\parskip=\medskipamount                 
\thicklines                         

\newcommand{\be}{\begin{equation}}
\newcommand{\ee}{\end{equation}}
\newcommand{\nbe}{\begin{equation*}}
\newcommand{\nee}{\end{equation*}}

\begin{document}

\thispagestyle{empty}

{\hbox to\hsize{
\vbox{\noindent April 2006 \hfill hep-th/0602115}}}
\noindent
\vskip1.3cm
\begin{center}

{\Large\bf $C$-deformation of Supergravity~\footnote{
Supported in part by the Japanese Society for Promotion of Science (JSPS)}}
\vglue.2in

Tomoya Hatanaka~\footnote{Email address: hatanaka-tomoya@c.metro-u.ac.jp}
and Sergei V. Ketov~\footnote{Email address: ketov@phys.metro-u.ac.jp}

{\it Department of Physics\\
     Tokyo Metropolitan University\\
     1--1 Minami-osawa, Hachioji-shi\\
     Tokyo 192--0397, Japan}
\end{center}
\vglue.2in
\begin{center}
{\Large\bf Abstract}
\end{center}

\noindent
A four-dimensional supergravity toy model in an arbitrary self-dual 
gravi-photon background is constructed in Euclidean space, by freezing out the
 gravi-photon field strength in the standard $N=(1,1)$ extended supergravity 
with two non-chiral gravitini. Our model has local $N=(1/2,0)$ supersymmetry. 
Consistency of the model requires the background gravi-photon field strength 
to be equal to the self-dual (bilinear) anti-chiral gravitino condensate.
 
\newpage

\section{Introduction}

As was shown by Ooguri and Vafa in ref.~\cite{ov}, the superworldvolume of a 
supersymmetric D-brane in a constant Ramond-Ramond type flux gives rise to
the remarkable new structure in the corresponding superspace, which is now
called {\it Non-AntiCommutativity} (NAC). The non-anticommutativity means that 
the fermionic superspace coordinates are no longer Grassmann (i.e. they no 
longer anti-commute), but satisfy a Clifford algebra. In other words, the 
impact of the RR flux on the D-brane dynamics can be simply described by the 
non-anticommutativity in the D-brane superworldvolume. In its turn, the 
non-anticommutativity in superspace can be easily described by the 
(Moyal-Weyl type) non-anticommutative star product among superfields, which 
gives rise to the NAC deformed supersymmetric field theories with partially 
broken supersymmetry \cite{fl,kpt,sei}. When gluino background is added, one 
can deform the anticommutation relation of the spinors in the D-brane 
worldvolume in order to recover full supersymmetry \cite{ov}.

As regards a D3-brane with its four-dimensional worldvolume, a ten-dimensional
(self-dual) five-form flux upon compactification to four dimensions gives rise
to the (self-dual) gravi-photon flux \cite{ov}. All recent studies of the NAC
supersymmetric field theories after the pioneering papers \cite{fl,kpt,sei}
were limited to rigid supersymmetry, i.e. without gravity. In this paper we
would like to investigate the impact of a self-dual gravi-photon flux on 
supergravity. 

The simplest supergravity model with a gravi-photon is the pure $N=(1,1)$ (or
$N=2$ in the Lorentz case) supergravity unifying gravity with electromagnetism.
Therefore, the easiest thing to do is to `freeze out' the gravi-photon field 
in that supergravity model to some self-dual value of its field strength. Of 
course, such a condition would break $N=(1,1)$ supersymmetry, so we should like
to investigate the residual supersymmetry, if any, and then cut the theory
properly, in our search for a supergravity model with lower local 
supersymmetry but with a non-vanishing self-dual gravi-photon background. It is
the task that we pursue in this paper. The Euclidean signature appears to be 
crucial here, similarly to the rigid NAC supersymmetric field theory 
\cite{kpt,sei}. We use the component formulation of supergravity 
\cite{nieu,fn}, as regards rigid N=1/2 supersymmetric field theories in 
superspace, see e.g., refs.~\cite{fl,kpt,sei,nac}.

Our paper is organized as follows. In sect.~2 we briefly introduce our 
notation. In sect.~3 we formulate the standard (Euclidean) four-dimensional 
$N=(1,1)$ supergravity in our notation. Sect.~4 is devoted to our results. Our
conclusion is sect.~5.

\section{About our notation}

Our notation is based on the standard review about supergravity \cite{nieu} 
with the four-dimensional spacetime signature $(+,+,+,+)$. We use lower case 
greek letters for (curved space) vector indices, $\m,\n,\ldots=1,2,3,4$, and
early lower case latin indices for (target space) vector indices, 
$a,b,\ldots=1,2,3,4$, early Capital latin letters for (anti)chiral spinor 
indices 
(dotted or undotted), $A,B,\ldots=1,2$, and middle lower case latin letters 
for indices of the $O(2)$ internal symmetry rotating two gravitini, 
$i,j,\ldots=1,2$. 

Raising and lowering of spinor indices is performed with the help of 
two-dimensional Levi-Civita symbols,
\be
\j\sb{A} = \j\sp{B} \e\sb{BA}, 
\quad 
\j\sp{A} = \e\sp{AB}\j\sb{B},
\qquad
\bar{\j}_{\dot{A}} = \bar{\j}^{\dot{B}}\e_{\dot{B}\dot{A}},
\quad
\bar{\j}^{\dot{A}} = \e^{\dot{A}\dot{B}}\bar{\j}_{\dot{B}}~,
\ee
while we have
\be
\e_{AB}=-\e_{BA},
\quad
\e^{AB}\e_{BC}=-\d\ud{A}{C},
\quad
\e_{12}=\e^{12}=-\e_{\dot{1}\dot{2}}=-\e^{\dot{1}\dot{2}}=1~.
\ee
Chiral and anti-chiral (with bars) spinors are independent in Euclidean space
with the signature $(4,0)$, as well as in Atiyah-Ward space with the signature
 $(2,2)$ \cite{kpt,kgn}.~\footnote{Chiral and anti-chiral spinors are 
related by complex conjugation in Minkowski space-time.} As a rule, we omit 
contracted spinor indices for simplicity in our equations, by using the 
notation
\be\eqalign{
\c\j~\equiv~& \c^{A}\j_{A}  =-\j_{A}\c^{A}=\j^{A}\c_{A}=\j\c
\cr
\bar{\c}\bar{\j}~\equiv~& \bar{\c}^{\dot{A}}\bar{\j}_{\dot{A}}
=-\bar{\j}_{\dot{A}}\bar{\c}^{\dot{A}}
=\bar{\j}^{\dot{A}}\bar{\c}_{\dot{A}}=\bar{\j}\bar{\c}~.\cr}
\ee
As regards, the (anti)self-dual parts of an antisymmetric tensor $K_{\m\n}\/$, 
we define
\be
K^{\pm}_{\m\n}\equiv
\ha \left(K_{\m\n}\mp\ha e\e_{\m\n\r\s}K^{\r\s}\right),
\qquad
\e_{1234}=1~,\qquad e=\det e^a_{\m}~.
\quad
\ee
The sigma-matrices in our notation are given by
\be
(\s)\du{a}{A\dot{B}}=(\vec{\s},iI)^{A\dot{B}},
\quad
(\s_a)_{\dot{A}B}=(\vec{\s},-iI)_{\dot{A}B}~,
\ee
while their triple (totally antisymmetric) product is given by
\be
\s^{abc A\dot{D}}\equiv
\fracmm{1}{6}\s^{A\dot{B}[a}\s^{b}_{\dot{B}{C}}\s^{c]C\dot{D}}~,
\ee
where $\vec{\s}$ are Pauli matrices and $I$ is unit matrix. 

Flat and curved vector indices are related by a vierbein $e^a_{\m}$ and 
 its inverse $e^{\m}_a$, as usual, e.g., $\s_{\m}=e^a_{\m}\s_a$ and  
$\s_{a}=e_a^{\m}\s_{\m}\/$, etc.

\section{$N=(1,1)$ supergravity}

Our starting point is the pure $N=(1,1)$ extended  supergravity in four 
Euclidean dimensions. The $N=(1,1)$ supergravity multiplet unifies a graviton 
field $e^a_{\m}$, two non-chiral gravitino fields $\j^i_{\m}\/$, and a 
gravi-photon gauge field  $A_{\m}\/$. The  $N=(1,1)$ supergravity action was 
first constructed in ref.~\cite{fn} by Noether procedure. When using our 
notation with chiral and antichiral gravitini, and making the $O(2)$ internal 
symmetry manifest, the $N=(1,1)$ supergravity Lagrangian of ref.~\cite{fn} 
reads 
\be \label{action}
\eqalign{
	\cl
=
&	-\fracmm{e}{2\k^2} R(e,\o)
	-ie\j^i_\m \s^{\m\r\s}D_\r(\o)\bar{\j}^i_\s
	-\fracmm{e}{4}F_{\m\n}F^{\m\n}
\cr
&	-\fracmm{e\k}{2\sqrt{2}}(\j^i_\m \j^j_\n)\e^{ij}
 	\left[F^{\m\n}+\hat{F}^{\m\n} \right]^{-}
 	-\fracmm{e\k}{2\sqrt{2}}(\bar{\j}^i_\m \bar{\j}^j_\n)\e^{ij}
 	\left[F^{\m\n}+\hat{F}^{\m\n} \right]^{+}~~,
\cr}\ee
where we have used the standard definitions \cite{nieu}, 
\be \label{defs}
F_{\m\n}=\pa_\m A_\n-\pa_\n A_\m\/,
\qquad
\hat{F}_{\m\n}
=
F_{\m\n}+\fracmm{\k}{\sqrt{2}}
(\j^i_\m \j^j_\n+\bar{\j}^i_\m \bar{\j}^j_\n)\e^{ij}~~,
\ee
of the gravi-photon field strength $F_{\m\n}$ and its supercovariant
extension $\hat{F}_{\m\n}$. The dimensional parameter $\kappa$ is the 
gravitational coupling constant. 

The spinor covariant derivative $D_{\m}(\o)\equiv D_{\m}$ contains the 
independent 
spin connection $\o^{ab}_{\m}$ that is supposed to be fixed as a function of 
the vierbein and gravitini by solving its algebraic equation of motion, 
$\d S/\d\o=0$ (it is known as the 1.5 order or Palatini formalism), as usual
\cite{nieu}.

By construction, the action $S=\int d^4 x\, \cl$ of eq.~(\ref{action}) 
is invariant under the following transformation rules of local $N=(1,1)$ 
supersymmetry:
\be \label{susytr}
\eqalign{
	\d e^a_\m 
&~=
	-\fracmm{i\k}{2}\left(\bar{\e}^i\s^a\j^i_\m+
\e^i\s^a\bar{\j}^i_\m\right),
\qquad
	\d A_{\m}
=
	-\fracmm{1}{\sqrt{2}}
	\e^{ij}\left(\e^i\j^j_\m+\bar{\e}^i\bar{\j}^j_\m\right),
\cr
	\d \j^i_\m
&~=
	\fracmm{1}{\k}D_\m\e^i
	-\fracmm{i}{\sqrt{2}}\e^{ij}\hat{F}^{+}_{\m\n}\s^\l\bar{\e}^j,
\qquad
	\d \bar{\j}^i_\m
=
	\fracmm{1}{\k}D_\m\bar{\e}^i
	+\fracmm{i}{\sqrt{2}}\e^{ij}\hat{F}^{-}_{\m\n}\s^\l\e^j,
\cr
\d F_{\m\n}
&~=
	-\fracmm{1}{\sqrt{2}}\e^{ij}
	\left[D_{[\m}(\e^i\j^j_{\n]})+D_{[\m}(\bar{\e}^i\bar{\j}^j_{\n]})
\right]~~,
\cr}
\ee
where $\e^i$ and $\bar{\e}^i$ stand for the infinitesimal chiral and 
anti-chiral anticommuting spinor parameters of local $N=(1,1)$ supersymmetry,
respectively. 

The supercovariant gravi-photon field strength $\hat{F}_{\m\n}$ transforms
covariantly under the transformations (\ref{susytr}) by construction (i.e. 
without derivatives of the supersymmetry parameters),
\be\label{supcurl}
	\d \hat{F}_{\m\n}=
	-\fracmm{1}{\sqrt{2}}
	\e^{ij}\left[\e^iD_{[\m}\j^j_{\n]}
	+\bar{\e}^iD_{[\m}\bar{\j}^j_{\n]}\right]
	+\fracmm{i\k}{2}
	\hat{F}^{+}_{\l[\m}\left(\bar{\e}^i\s^\l\j^i_{\n]}\right)
	+\fracmm{i\k}{2}
	\hat{F}^{-}_{\l[\m}\left(\e^i\s^\l\bar{\j}^i_{\n]}\right)~.
\ee

\section{$C$-deformation}

We now impose the condition
\be\label{cond} 	
\hat{F}^{+}_{\m\n}= C_{\m\n}(x), \qquad
	\hat{F}^{-}_{\m\n}= 0~,\ee
where $\hat{F}_{\m\n}$ is the covariantized gravi-photon field strength
(\ref{defs}), and $C_{\m\n}(x)=C^+_{\m\n}(x)$ is an arbitrary given function.
We have chosen  $\hat{F}_{\m\n}$ instead of $F_{\m\n}$ in eq.~(\ref{cond}) 
because $\hat{F}_{\m\n}$ is a tensor under the local supersymmetry (sect.~3),
though it turns to be unimportant (see the end of this section). Also, we do
not require the background $C_{\m\n}(x)$ to be constant, because it turns 
out to be unessential too (cf. a coordinate-dependent deformation of rigid 
$N=1/2$ supersymmetry in superspace \cite{ass}). 

Of course, the condition (\ref{cond}) is not compatible with the full $N=(1,1)$
local supersymmetry (\ref{susytr}). The consistency condition 
\be \label{cons}
 \d \hat{F}_{\m\n}=0~, \ee
has, however, the residual $N=(1/2,0)$ local supersymmetry with the
infinitesimal spinor parameter $\e^1(x)$, when choosing
\be\label{res}
\e^2=\bar{\e}^1=\bar{\e}^2=0\qquad{\rm and}\qquad \j^2_\m=0~.\ee
Note that then $\d\j^2_\m=0$ is automatically satisfied, while the Lagrangian
(\ref{action}) takes the form
\be \label{clag}
\eqalign{
	\cl
	=
&
	-\fracmm{e}{2\k^2} R(e,\o)
	-ie\j^1_\m \s^{\m\r\s}D_\r(\o)\bar{\j}^{1}_\s
	-\fracmm{e}{4}C_{\m\n}C^{\m\n}
\cr
&	
	-\fracmm{e}{2}
	\left[C_{\m\n}+\fracmm{e}{4}\e_{\m\n\r\s}\fracmm{\k}{\sqrt{2}}
	\left(\bar{\j}^{i\r} \bar{\j}^{j\s}\right)\e^{ij}\right]
	\fracmm{\k}{\sqrt{2}}\left(\bar{\j}^{k\m} \bar{\j}^{l\n}\right)\e^{kl}
~.\cr}\ee
The field $\bar{\j}^{2}_{\g\dot{B}}$ enters the Lagrangian (\ref{clag}) as a 
Lagrange multiplier, so that it gives rise to a constraint (after varying the
action with respect to that field),
\be \label{constr}
\left[C_{\m\n}
+\fracmm{e}{2}\e_{\m\n\r\s}\fracmm{\k}{\sqrt{2}}
\left(\bar{\j}^{i\r} \bar{\j}^{j\s}\right)\e^{ij}\right]
g^{\n\g}\bar{\j}^{1\m\dot{B}}
=0~,\ee
whose solution is
\be \label{csol}
 C_{\m\n}=
-\fracmm{e}{2}\e_{\m\n\r\s}\fracmm{\k}{\sqrt{2}}
\left(\bar{\j}^{i\r} \bar{\j}^{j\s}\right)\e^{ij}
=
\fracmm{\k}{\sqrt{2}}
\left(\bar{\j}^i_\m \bar{\j}^j_\n\right)\e^{ij}~.
\ee

In the simplified notation
\be \label{simple}
\j_\m\equiv\j^1_\m,\quad
\bar{\j}_\m\equiv\bar\j^{1}_\m,
\quad
\bar{\c}_\m\equiv\bar\j^{2}_\m,
\ee
we thus arrive at the Lagrangian
\be \label{1/2lag}
\cl=
-\fracmm{e}{2\k^2} R(e,\o)
-ie\j_\m \s^{\m\r\s}D_\r(\o)\bar{\j}_\s
-\fracmm{e}{2}C_{\m\n}C^{\m\n}~,
\ee
where 
\be \label{+constr}
C_{\m\n}=\fracmm{\k}{\sqrt{2}}\left[\bar{\j}_{[\m}\bar{\c}_{\n]}\right]^{+}~.
\ee
By our construction, the supergravity action $S=\int d^4 x\, \cl$ with the
Lagrangian (\ref{1/2lag}) is invariant under the $N=1/2$ local supersymmetry 
with the transformation laws 
\be\label{1/2susy}
	\d e^a_\m=-\fracmm{i\k}{2}\e\s^a\bar{\j}_\m~,
\quad
	\d \j_\m=\fracmm{1}{\k}D_\m\e~,
\quad
	\d \bar{\j}_\m=0~,
\quad
	\d \bar{\c}_\m=0~.
\ee
Note that eqs.~(\ref{defs}) and (\ref{csol}) also imply
\be \label{no1}
F_{\m\n}=0~.\ee

The result (\ref{no1}) may prompt us to choose another constraint,
\be \label{cond2}
F^{+}_{\m\n}=C_{\m\n}(x), 	\qquad
	F^{-}_{\m\n}= 0~,\ee
{}from the very beginning of this section, instead of eq.~(\ref{cond}). Then 
its consistency with local supersymmetry, 
\be \d F_{\m\n}=0~,\ee
under the transformations (\ref{susytr}) again has a solution (\ref{res}),
with an arbitrary infinitesimal parameter $\e^1(x)$. The Lagrangian 
(\ref{action}) then takes the form
\be 
\eqalign{
	\cl
	=
&
	-\fracmm{e}{2\k^2} R(e,\o)
	-ie\j^1_\m \s^{\m\r\s}D_\r(\o)\bar{\j}^{1}_\s
	-\fracmm{e}{4}C_{\m\n}C^{\m\n}
\cr
&	
	-\fracmm{e}{2}
	\left[
		2C_{\m\n}
		+
		\left(
			\fracmm{\k}{\sqrt{2}}
			\left(\bar{\j}^i_\m \bar{\j}^j_\n\right)\e^{ij}
		\right)^{+}
	\right]
	\fracmm{\k}{\sqrt{2}}
	\left(\bar{\j}^{k\m} \bar{\j}^{l\n}\right)\e^{kl}~.
\cr}\ee
The algebraic equation of motion of the field $\bar{\j}^{2}_{\g\dot{B}}$,
\be 
\left[
	C_{\m\n}
	+
	\left(
		\fracmm{\k}{\sqrt{2}}
		\left(\bar{\j}^i_\m \bar{\j}^j_\n\right)\e^{ij}
	\right)^{+}
\right]
g^{\n\g}\bar{\j}^{1\m\dot{B}}
=0~,
\ee
now has a solution
\be \label{con4}
C_{\m\n}=
-\fracmm{\k}{\sqrt{2}}
\left(\bar{\j}^i_\m \bar{\j}^j_\n\right)\e^{ij}
=
\fracmm{e}{2}\e_{\m\n\r\s}\fracmm{\k}{\sqrt{2}}
\left(\bar{\j}^{i\r} \bar{\j}^{j\s}\right)\e^{ij}~.
\ee
When using the notation (\ref{simple}), we arrive at the $N=1/2$ supergravity
 Lagrangian 
\be \label{lag2c}
\cl =
-\fracmm{e}{2\k^2} R(e,\o)
-ie\j_\m \s^{\m\r\s}D_\r(\o)\bar{\j}_\s
+\fracmm{e}{4}C_{\m\n}C^{\m\n}
\ee
subject to the constraint
\be \label{csol2}
 C_{\m\n}=
-\fracmm{\k}{\sqrt{2}}
\left[\bar{\j}_{[\m}\bar{\c}_{\n]}\right]^{+}~,
\ee
which are both invariant under the same $N=1/2$ local supersymmetry 
transformations (\ref{1/2susy}). In this case we have
\be\label{no2} \hat{F}_{\m\n}=0\ee
instead of eq.~(\ref{no1}).

\section{Conclusion}

The new model we constructed is, of course, a toy model with local $N=1/2$
supersymmetry. Nevertheless, we found that an $N=1/2$ supergravity is possible,
while it can be very simple, like the $C$-deformed $N=1/2$ supersymmetric 
gauge theory constructed in the NAC-deformed superspace \cite{sei}. Perhaps, 
the most remarkable feature of our construction  is the very simple relation
it implies between the expectation value of $C^+_{\m\n}$ and that of the 
self-dual product of two anti-chiral gravitini --- see eqs.~(\ref{+constr}) and
(\ref{csol2}). Another approach to a construction of a $C$-deformed $N=1/2$ 
supergravity in four Euclidean dimensions can be based on the NAC deformed 
$N=(1/2,1/2)$ superspace, by imposing the NAC relation on the chiral superspace
coordinates, $\{\q^{A},\q^{B}\}_*=C^{AB}$, with $C^{\m\n}=
C^{AB}\e_{BC}(\s^{\m\n})\du{A}{C}$. When using the Moyal-Weyl-type star 
product $(*)$ for the supergravity superfields \cite{sei} and applying the 
explicit superfield star-product summation formulae of ref.~\cite{nac}, it may
be possible to get another $C$-deformed $N=1/2$ supergravity action in
components (in a Wess-Zumino gauge). However, the last procedure seems to be 
much more complicated~\cite{nat}, and we do not claim that the result is going
to be equivalent to our model constructed above. 

A non-vanishing gravitino condensate is the well known tool for spontaneous
breaking of local supersymmetry, which may lead to a natural solution to the
hierarchy problem in elementary particles physics (see e.g., the pioneering
paper \cite{witten} for the dynamical mechanism due to gravitational instantons
 in quantum gravity, and ref.~\cite{kita} for its possible realization
in superstring theory). Because of the relations (19) and (28), the 
$C$-deformation may be the viable alternative mechanism of spontaneous local
supersymmetry breaking. To explore its physical consequences, one has to go
beyond our toy model of $C$-deformed pure supergravity by adding some matter 
content \cite{nat}.

\end{document}